\documentclass[amsmath, amssymb, aps, physrev, 10pt]{revtex4-2}

\usepackage{graphicx}
\usepackage{dcolumn}
\usepackage{bm}
\usepackage{dsfont}
\usepackage{amsmath}
\usepackage{color}
\usepackage{MnSymbol}
\usepackage{subfigure}
\usepackage{ragged2e}
\usepackage{float}
\usepackage{booktabs}
\usepackage{silence}
\usepackage{tikz}
\usetikzlibrary{decorations.pathreplacing}
\usepackage{lineno}

\def\M11{$\mathbb M^{1+1}$}

\def\precstar{\prec\hspace{-1.4mm}\star\hspace{0.8mm}}
\def\CCC{\mathbf{C}}
\def\LLL{\mathbf{L}}

\def\cR{\mathcal R}

\newcommand{\AdS}{\mathrm{AdS}}
\newcommand{\dS}{\mathrm{dS}}
\newcommand{\Mink}{\mathbb{M}}
\newcommand{\dd}{\mathrm{d}}
\newcommand{\Kcs}{\mathcal{K}}
\newcommand{\ThetaF}{\Theta_{+}}
\newcommand{\avg}[1]{\left\langle #1\right\rangle}

\begin{document}

\title{\textbf{Construction of Causal Set Propagators in $2+1$-Dimensional Constant-Curvature Spacetimes from Two-Link Paths}}
\author{Arsim Kastrati}
\email{Contact author: arsim.kastrati@uni-wuerzburg.de}
\affiliation{Faculty for Physics and Astronomy, Julius Maximilians University Würzburg, Am Hubland, 97074 Würzburg, Germany}
\date{\today}


\begin{abstract}
We study a link-based retarded scalar propagator on causal sets sprinkled into $2+1$-dimensional Minkowski, anti-de Sitter, and de Sitter spacetimes. The construction is based on the squared link matrix~$\mathbf{L}^2$, whose expectation value counts two-link paths between fixed endpoints. In the high-density limit, the dominant contribution localizes near the intersection of the null boundaries of the Alexandrov interval, and the flat-space normalization remains unchanged in constant curvature. The resulting kernels agree with the corresponding continuum Klein–Gordon propagators for a nominally massless scalar with effective curvature coupling $\xi_{\rm link}=1/6$. Since the conformal coupling in $2+1$ spacetime dimensions is $1/8$, this value is not a conformal-coupling effect. Together with link-based results in $1+1$ and $3+1$ dimensions, where the same coefficient $1/6$ appears, this suggests that the coupling may reflect an ultraviolet property of the link structure, reminiscent of the combination $(1/6-\xi)\mathcal{R}$ in local heat-kernel expansions.
\end{abstract}

\maketitle
\parskip 2mm

\section{Introduction}
\label{sec:introduction}

Causal set theory replaces a Lorentzian spacetime by a locally finite partially ordered set of events, whose order encodes causal precedence and its cardinality encodes spacetime volume ~\cite{Bombelli1987,Malament1977,HawkingKingMcCarthy1976,Surya2019}.  Because retarded Green functions are intrinsically causal objects, scalar propagation provides a particularly direct setting in which to compare continuum and discrete structures.  Several complementary constructions have been developed. Most prominently, Johnston introduced a hop-and-stop path sum in which causal trajectories are weighted and summed by matrix methods, whose mean approaches the corresponding continuum retarded propagators for sprinklings into $1+1$- and $3+1$-dimensional Minkowski, de Sitter and anti-de Sitter spacetimes ~\cite{Johnston2008,Johnston2010,KastratiHinrichsen2026AdS,AhmedDowkerSurya2017}. The retarded kernel also underlies covariant causal-set quantum field constructions such as the causal-set Feynman propagator and the Sorkin--Johnston state ~\cite{Surya:2018byh,Johnston2009,Afshordi2012}.

A causal set $\mathcal{C}$ is a locally finite partially ordered set
whose order relation $\prec$ encodes causal precedence, while the
number of elements provides a discrete measure of spacetime volume.
Its causal structure is described by the causal matrix $\CCC$, which
records all causal relations
\begin{equation}
    \CCC_{xy} =
    \begin{cases}
        1, & y\prec x,\\
        0, & \text{otherwise},
    \end{cases}
\end{equation}
and the link matrix $\LLL$, which describes only irreducible ones
\begin{equation}
    \LLL_{xy} =
    \begin{cases}
        1, & y\precstar x,\\
        0, & \text{otherwise}.
    \end{cases}
\end{equation}
Here $y\precstar x$ denotes a link, meaning that no element lies
causally between $y$ and $x$. Thus, $\CCC$ is the transitive closure
of the link relation, whereas $\LLL$ is the adjacency matrix of
the partial ordered set of events. In particular, $(\LLL^n)_{xy}$ counts paths
from $y$ to $x$ consisting of exactly $n$ links. Following earlier link-based constructions, we assume $\LLL$ to be the
central object from which the propagator is built. Unlike the
causal matrix, the link matrix is sensitive to the volume of the
Alexandrov interval between two events.

In $3+1$ dimensions the massless flat retarded Green function can be represented directly by a density-rescaled link matrix~\cite{Johnston2008}, and the same prescription was shown to reproduce the massless conformally coupled Green function in de Sitter (dS) spacetime and in a conformally flat globally hyperbolic patch of anti-de Sitter (AdS) spacetime ~\cite{AhmedDowkerSurya2017}.  More recently, we proposed a construction centered entirely on links in $1+1$ dimensions: the normalized exponential $\exp (\LLL)$ reproduces the massless retarded Minkowski propagator after averaging, and the massive theory follows from the usual scattering series ~\cite{HinrichsenKastrati2026}.  Its response to constant curvature gives, to leading order analytically and with numerical support in $\AdS_2$ and $\dS_2$, an effective curvature coupling $\xi=1/6$ ~\cite{Kastrati2026Curvature}.  Related numerical work has also tested causal-set path-sum propagators in $1+1$D de Sitter and anti-de Sitter spacetimes ~\cite{Aslanbeigi:2013fga,KastratiHinrichsen2026AdS}.

So far, the $2+1$ dimensional case received less attention.  Johnston discussed a candidate based on the relation between proper time and Alexandrov volume, while Nomaan X, Dowker, and Surya later proposed a candidate based on the inverse longest-chain length, motivated by the continuum approximation of longest chains to timelike distance ~\cite{Johnston2010,BrightwellGregory1991,AhmedDowkerSurya2017}.  These constructions are natural, but neither uses the link matrix as the sole elementary object in the same way the $1+1$- and $3+1$-dimensional link prescriptions do.

The purpose of the present paper is to fill that gap at the level of the mean retarded kernel.  We propose that the massless kernel $\Kcs^{(d)}_{0}(x,y)$ in $d=2+1$ dimensions is given by
\begin{equation}
    \Kcs^{(3)}_{0}(x,y)
    :=\mathcal{N}_3\rho^{1/3}\avg{(\LLL^2)_{xy}},\quad \mathcal{N}_3=\left[\frac{3}{2\pi\Gamma(2/3)}
    \left(\frac{\pi}{12}\right)^{2/3}\right]^2,
    \label{eq:def_kernel_intro}
\end{equation}
with the normalization $\mathcal{N}_3$ fixed in flat spacetime.  The central questions are whether the same definition and normalization survives in constant curvature and which continuum Klein--Gordon operator is selected by the bare link-square kernel.

In the high-density limit, the expectation value of $\LLL^2$ is dominated by intermediate points near the intersection of the corners of the spanned causal diamond. Since the corresponding small subintervals have the universal leading Minkowski volume, curvature does not modify the normalization $\mathcal{N}_3$, but enters through the finite-separation dependence of the propagator. As a main result of this paper, for timelike separation $\tau$, we obtain
\begin{align}
    \Kcs^{\Mink_3}_{0}(\tau)
    &=\frac{1}{2\pi\tau},
    \label{eq:main_M}\\
    \Kcs^{\AdS_3}_{0}(\tau)
    &=\frac{1}{2\pi\ell\sin(\tau/\ell)},
    \label{eq:main_AdS}\\
    \Kcs^{\dS_3}_{0}(\tau)
    &=\frac{1}{2\pi\ell\sinh(\tau/\ell)}.
    \label{eq:main_dS}
\end{align}
Here $\ell$ denotes the curvature radius, and both curved expressions
reduce to the Minkowski result in the flat limit. Comparison with the continuum Klein--Gordon equation,
\begin{equation}
    \left(\Box_g-m^2-\xi\cR\right)G_{\rm ret}
    =\delta_g,
    \label{eq:KG_convention}
\end{equation}
shows that the curved kernels correspond to a nominally massless scalar
with an effective curvature coupling
\begin{equation}
   \xi_{\rm link}=\frac16.
    \label{eq:xi_main}
\end{equation}
They therefore differ from the minimally coupled massless propagators
on the same backgrounds, as also supported by our numerical results. The appearance of the same coupling in link-based constructions in
$1+1$ and $3+1$ dimensions is noteworthy. Since the conformal curvature coupling in $2+1$ dimensions is $\xi_{\rm conf}=1/8$, the value $\xi_{\rm link}=1/6$ cannot be attributed to conformal coupling. The dimension-independent combination $(1/6-\xi)\cR$
in local heat-kernel expansions provides a suggestive continuum
analogy, although a common microscopic explanation remains to be
established.

The paper is organized as follows. Sec. \ref{sec:link_square} introduces the link-square kernel and establishes its flat-space normalization, before deriving the high-density limits in $\mathrm{AdS}_3$ and $\mathrm{dS}_3$. These results are then compared with the corresponding continuum Klein--Gordon propagators and supported by numerical simulations. The section concludes by discussing the relation to link-based constructions in other spacetime dimensions and the possible ultraviolet origin of the effective curvature coupling. Sec. \ref{sec:outlook} summarizes the results and outlines directions for further work. Technical details of the asymptotic calculations and the exact Alexandrov interval volumes are provided in the appendices.

\section{The flat link-square kernel}
\label{sec:link_square}

Let $\mathcal C$ be a causal set obtained by a Poisson sprinkling of events with density $\rho$ into a Lorentzian spacetime $(M,g)$.  For two events $y\prec x$ that are held fixed as endpoints, the probability that they form a link is the probability that the Alexandrov interval $I(y,x)$ contains no sprinkled point.  Thus
\begin{equation}
    \avg{\LLL(x,y)}
    =\ThetaF(x,y)\exp[-\rho V(x,y)],
    \label{eq:mean_link}
\end{equation}
where $V(x,y)$ is the spacetime volume of $I(y,x)$ and $\ThetaF$ restricts the kernel to the causal future.  The dependence on $V$ is the essential source of both the flat-space localization and the curvature sensitivity of a link.

The square of the link matrix contains one intermediate element,
\begin{equation}
    (\LLL^2)_{xy}=\sum_{z\in \mathcal C}\LLL_{xz}\LLL_{zy}.
    \label{eq:L2sum}
\end{equation}
For $y\prec z\prec x$, the open intervals $I(y,z)$ and $I(z,x)$ are disjoint.  Their Poisson emptiness probabilities therefore factorize and the exact mean is given by
\begin{equation}    
    F_2(x,y)
    :=\avg{(\LLL^2)_{xy}}
    =\rho\int_{z \in I(y,x)}\dd V_g(z)\,
    e^{-\rho V(x,z)}e^{-\rho V(z,y)} .    
    \label{eq:F2general}
\end{equation}
The factor $\rho$ reflects the one independent intermediate matrix sum, since the expected number of sprinkled elements in $dV_g(z)$ is $\rho\,dV_g(z)$.  Eq.~\eqref{eq:F2general} is the starting point for all three backgrounds considered below.

\subsection{Minkowski spacetime}
\label{sec: Mink}
In $2+1$-dimensional Minkowski spacetime, we use polar coordinates with the metric
\begin{equation}
    \dd s^2=-\dd t^2+\dd r^2+r^2\dd\phi^2,
    \qquad \dd V=r\,\dd t\,\dd r\,\dd\phi.
    \label{eq:flatpolar}
\end{equation}
Without loss of generality, we choose the two points
\begin{equation}
    p=(-a,0),\qquad q=(a,0),\qquad \tau=2a.
    \label{eq:flatendpoints}
\end{equation}
The volume of a causal diamond in terms of proper time is
\begin{equation}
    V_{\Mink_3}(\tau)=\frac{\pi}{12}\tau^3.
    \label{eq:flat_volume}
\end{equation}
\begin{figure}[t]
\centering
\includegraphics[width=30mm]{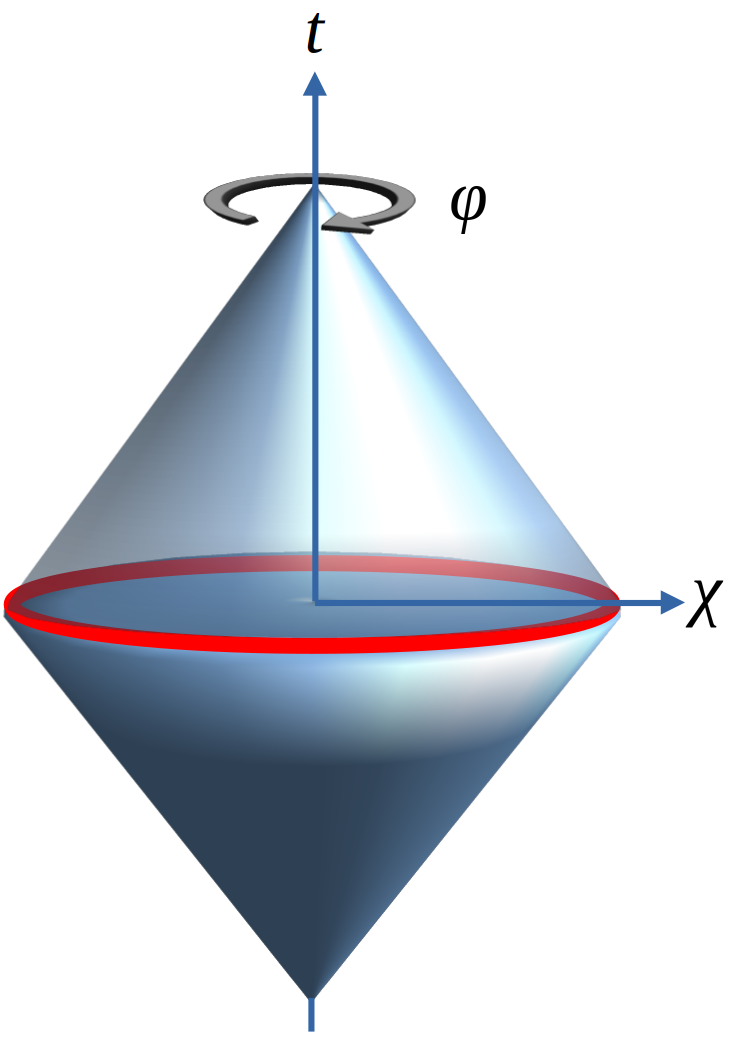}
\caption{Causal diamond defined by two timelike-separated events in $2+1$ dimensions. The future light cone of the earlier event and the past light cone of the later event intersect along the circular waist shown in red. The coordinates $t$, $\chi$, and $\phi$ denote time, radial position, and azimuthal angle. In the high-density limit, the dominant contribution to the link-square expectation value comes from intermediate points concentrated near this waist.}
\label{fig:null-waist-region}
\end{figure}
For the fixed timelike-separated endpoints \eqref{eq:flatendpoints}, the exponential in Eq.~\eqref{eq:F2general} localizes the intermediate point near the circular waist at which the future light cone of the earlier endpoint meets the past lightcone of the later endpoint, as shown in Fig. \ref{fig:null-waist-region}.  A boundary-layer evaluation, given in Appendix~\ref{app:flat}, yields
\begin{equation}
    F_2^{\Mink_3}(\tau)
    \sim
    \frac{2\pi}{9}\Gamma\!\left(\frac23\right)^2
    \left(\frac{\pi}{12}\right)^{-4/3}\frac{\rho^{-1/3}}{\tau}
    \qquad (\rho\to\infty).
    \label{eq:flat_F2_asymptotic}
\end{equation}
It follows that the normalization
\begin{equation}    
    \mathcal{N}_3=\left[\frac{3}{2\pi\Gamma(2/3)}
    \left(\frac{\pi}{12}\right)^{2/3}\right]^2
    \label{eq:A3}
\end{equation}
gives
\begin{equation}   
    \lim_{\rho\to\infty}\mathcal{N}_3\rho^{1/3}
    F_2^{\Mink_3}(\tau)
    =\frac{1}{2\pi\tau}.    
    \label{eq:flat_limit}
\end{equation}
With retarded support restored, the right-hand side is the massless retarded scalar Green function in $2+1$-dimensional Minkowski spacetime,
\begin{equation}
    G^{\Mink_3}_{0,\rm ret}(x,y)
    =\frac{\ThetaF(x,y)}{2\pi\tau(x,y)}.
    \label{eq:flat_green}
\end{equation}

\subsection{Anti-de Sitter spacetime}
\label{sec:AdS}

We treat $\AdS_3$ in global conformal coordinates,
\begin{equation}
    \dd s^2=\frac{\ell^2}{\cos^2\chi}
    \left(-\dd t^2+\dd\chi^2+\sin^2\chi\,\dd\phi^2\right),
    \qquad t\in\mathbb R,\,\, 0\le\chi<\pi/2,\,\, 0\le\phi<2\pi,
    \label{eq:AdSmetric}
\end{equation}
where $\ell$ is the curvature radius. The volume element is
\begin{equation}
    \dd V=\ell^3\frac{\sin\chi}{\cos^3\chi}\,
    \dd t\,\dd\chi\,\dd\phi,
    \qquad
    \cR_{\AdS_3}=-\frac{6}{\ell^2},
    \label{eq:AdSmeasureR}
\end{equation}
By symmetry we place the endpoints on the central geodesic at
\begin{equation}
    p=(-a,0),\qquad q=(a,0),\qquad \tau=2\ell a,
    \qquad 0<a<\frac{\pi}{2}.
    \label{eq:AdSendpoints}
\end{equation}
The exact Alexandrov volume is
\begin{equation}    
    V_{\AdS_3}(\tau)
    =2\pi\ell^3\left[
       \tan\left(\frac{\tau}{2\ell}\right)
       -\frac{\tau}{2\ell}
    \right].
    \label{eq:AdSvolume}
\end{equation}
At short distance,
\begin{equation}
    V_{\AdS_3}(\tau)
    =\frac{\pi}{12}\tau^3
     +\frac{\pi}{120\ell^2}\tau^5
     +O\!\left(\frac{\tau^7}{\ell^4}\right).
    \label{eq:AdSvolsmall}
\end{equation}
The leading term is exactly Eq.~\eqref{eq:flat_volume}; curvature first enters at relative order $\tau^2/\ell^2$.  This local fact already suggests that the density power and the normalization $\mathcal{N}_3$ should remain unchanged.

At large $\rho$, Eq.~\eqref{eq:F2general} localizes near the waist $t=0$, $\chi=a$.  Introducing nonnegative distances from the two null boundaries,
\begin{equation}
    u=a-t-\chi,\qquad v=a+t-\chi,
    \label{eq:AdSuv}
\end{equation}
one finds to leading order
\begin{equation}
    \dd V\simeq
    \pi\ell^3\frac{\sin a}{\cos^3a}\,\dd u\,\dd v,
    \qquad
    V(q,z)\simeq c(a)u^{3/2},\qquad
    V(z,p)\simeq c(a)v^{3/2},
    \label{eq:AdSboundarydata}
\end{equation}
where
\begin{equation}
    c(a)=\frac{\pi}{12}\left(2\ell^2\tan a\right)^{3/2}.
    \label{eq:ca}
\end{equation}
The detailed derivation is given in Appendix~\ref{app:AdS}.  Substitution into Eq.~\eqref{eq:F2general} gives
\begin{equation}
    F_2^{\AdS_3}(\tau)
    \sim
    \frac{4\pi}{9}\Gamma\!\left(\frac23\right)^2
    \ell^3\frac{\sin a}{\cos^3a}\,c(a)^{-4/3}\rho^{-1/3}.
    \label{eq:AdSF2asymptotic}
\end{equation}
Using the normalization in Eq.~\eqref{eq:A3}, all Gamma-function and local-volume factors cancel and we obtain
\begin{align}
    \lim_{\rho\to\infty}\mathcal{N}_3\rho^{1/3}F_2^{\AdS_3}(\tau)
    &=\frac{1}{4\pi\ell\sin a\cos a}
    \nonumber\\
    &=\frac{1}{2\pi\ell\sin(\tau/\ell)}.
    \label{eq:AdSlimit}
\end{align}
Thus curvature does not renormalize the leading local coefficient $\mathcal{N}_3$; instead, it survives in the finite-separation dependence of the continuum kernel.  The flat limit is immediate:
\begin{equation}
    \ell\to\infty:\qquad
    \frac{1}{2\pi\ell\sin(\tau/\ell)}
    \longrightarrow\frac{1}{2\pi\tau}.
    \label{eq:AdSflatlimit}
\end{equation}

\subsection{de Sitter spacetime}
\label{sec:dS}

The corresponding calculation in $\dS_3$ is analogous, but it is useful to display the geometrical quantities that change sign under the passage from negative to positive curvature.  In global conformal coordinates we use
\begin{equation}
    \dd s^2=\frac{\ell^2}{\cos^2 t}
    \left(-\dd t^2+\dd\chi^2+\sin^2\chi\,\dd\phi^2\right),
    \qquad -\frac{\pi}{2}<t<\frac{\pi}{2}\,\, 0\le\chi<\pi/2,\,\, 0\le\phi<2\pi,
    \label{eq:dSmetric}
\end{equation}
with
\begin{equation}
    \dd V=\ell^3\frac{\sin\chi}{\cos^3 t}\,
    \dd t\,\dd\chi\,\dd\phi,
    \qquad
    \cR_{\dS_3}=+\frac{6}{\ell^2}.
    \label{eq:dSmeasureR}
\end{equation}
For symmetric endpoints $p=(-a,0)$ and $q=(a,0)$, the proper-time separation along the central geodesic is
\begin{equation}
    \tau=2\ell\operatorname{artanh}(\sin a).
    \label{eq:dStauofa}
\end{equation}
The exact Alexandrov volume is
\begin{equation}   
    V_{\dS_3}(\tau)
    =2\pi\ell^3\left[
       \frac{\tau}{2\ell}
       -\tanh\left(\frac{\tau}{2\ell}\right)
    \right],    
    \label{eq:dSvolume}
\end{equation}
so that
\begin{equation}
    V_{\dS_3}(\tau)
    =\frac{\pi}{12}\tau^3
     -\frac{\pi}{120\ell^2}\tau^5
     +O\!\left(\frac{\tau^7}{\ell^4}\right).
    \label{eq:dSvolsmall}
\end{equation}
The opposite sign of the first curvature correction is consistent with $\cR_{\dS_3}=-\cR_{\AdS_3}$ at fixed $\ell$.

The high-density integral again localizes at the waist $t=0$, $\chi=a$.  With the same variables as in Eq.~\eqref{eq:AdSuv}, the endpoint interval volumes have the same leading form,
\begin{equation}
    V(q,z)\simeq c(a)u^{3/2},\qquad
    V(z,p)\simeq c(a)v^{3/2},
    \label{eq:dSsmallV}
\end{equation}
with $c(a)$ given by Eq.~\eqref{eq:ca}, whereas the waist measure is now
\begin{equation}
    \dd V\simeq\pi\ell^3\sin a\,\dd u\,\dd v.
    \label{eq:dSwaistmeasure}
\end{equation}
Appendix~\ref{app:dS} gives the derivation.  Therefore
\begin{equation}
    F_2^{\dS_3}(\tau)
    \sim
    \frac{4\pi}{9}\Gamma\!\left(\frac23\right)^2
    \ell^3\sin a\,c(a)^{-4/3}\rho^{-1/3}.
    \label{eq:dSF2asymptotic}
\end{equation}
The same flat normalization gives
\begin{align}
    \lim_{\rho\to\infty}\mathcal{N}_3\rho^{1/3}F_2^{\dS_3}(\tau)
    &=\frac{1}{4\pi\ell}\frac{\cos^2a}{\sin a}
    \nonumber\\
    &=\frac{1}{2\pi\ell\sinh(\tau/\ell)},
    \label{eq:dSlimit}
\end{align}
where we used Eq.~\eqref{eq:dStauofa}.  Again,
\begin{equation}
    \ell\to\infty:\qquad
    \frac{1}{2\pi\ell\sinh(\tau/\ell)}
    \longrightarrow\frac{1}{2\pi\tau}.
    \label{eq:dSflatlimit}
\end{equation}
Eqs.~\eqref{eq:AdSlimit} and \eqref{eq:dSlimit} show that the same microscopic link-square normalization extends continuously to both signs of constant curvature.

\subsection{Continuum Klein--Gordon interpretation}
\label{sec:KG}

We now identify the continuum operator represented by the bare link-square kernel.  Let
\begin{equation}
    M^2:=m^2+\xi \cR.
    \label{eq:M2def}
\end{equation}
For geodesically invariant functions of timelike distance, the homogeneous part of Eq.~\eqref{eq:KG_convention} reduces in $\AdS_3$ to
\begin{equation}
    G''(\tau)+\frac{2}{\ell}\cot\left(\frac{\tau}{\ell}\right)G'(\tau)
    +M^2G(\tau)=0,
    \label{eq:AdSradialKG}
\end{equation}
and in $\dS_3$ to
\begin{equation}
    G''(\tau)+\frac{2}{\ell}\coth\left(\frac{\tau}{\ell}\right)G'(\tau)
    +M^2G(\tau)=0.
    \label{eq:dSradialKG}
\end{equation}
A direct retarded solution normalized to the flat short-distance singularity is, respectively,
\begin{align}
    G^{\AdS_3}_{m,\xi}(\tau)
    &=\frac{1}{2\pi\ell}
      \frac{\cos(\nu_-\tau/\ell)}{\sin(\tau/\ell)},
    &\nu_-^2&=1+\ell^2M^2,
    \label{eq:AdSKGsolution}\\
    G^{\dS_3}_{m,\xi}(\tau)
    &=\frac{1}{2\pi\ell}
      \frac{\cosh(\nu_+\tau/\ell)}{\sinh(\tau/\ell)},
    &\nu_+^2&=1-\ell^2M^2.
    \label{eq:dSKGsolution}
\end{align}
These expressions refer to the direct timelike Green function. The link-square limits \eqref{eq:AdSlimit} and \eqref{eq:dSlimit} contain no cosine or hyperbolic-cosine factor.  Thus $\nu_-=0$ in $\AdS_3$ and $\nu_+=0$ in $\dS_3$, which gives
\begin{equation}
    M^2_{\rm bare}=
    \begin{cases}
       -\ell^{-2}, & \AdS_3,\\[2mm]
       +\ell^{-2}, & \dS_3.
    \end{cases}
    \label{eq:Mbarepm}
\end{equation}
Since $\cR=\mp6/\ell^2$, both cases are summarized by
\begin{equation}
    M^2_{\rm bare}=\frac{\cR}{6}.
    \label{eq:MbareR6}
\end{equation}
For the nominally massless bare construction, $m=0$, Eq.~\eqref{eq:MbareR6} therefore implies
\begin{equation}
    \xi_{\rm link}=\frac16.
    \label{eq:xilink}
\end{equation}
This coupling has not been inserted into $\LLL$ or $\LLL^2$ by hand.  It is inferred from the continuum operator selected by the high-density link-square expectation.

It is useful to contrast the result with minimal coupling.  Setting $m=\xi=0$ in Eqs.~\eqref{eq:AdSKGsolution} and \eqref{eq:dSKGsolution} gives
\begin{align}
    G^{\AdS_3}_{0,0}(\tau)
    &=\frac{1}{2\pi\ell}\cot\left(\frac{\tau}{\ell}\right),
    \label{eq:AdSminimal}\\
    G^{\dS_3}_{0,0}(\tau)
    &=\frac{1}{2\pi\ell}\coth\left(\frac{\tau}{\ell}\right).
    \label{eq:dSminimal}
\end{align}
These differ from the bare link-square kernels.  Thus the latter should not be described as minimally coupled massless propagators with a finite-density artifact.  Their curvature scale remains finite when $\rho\to\infty$ and disappears only in the flat limit $\ell\to\infty$.

For a target minimally coupled scalar of physical mass $m$, the required change in the Klein--Gordon parameter is
\begin{equation}
    \Delta M^2=M^2_{\rm target}-M^2_{\rm bare}
    =m^2-\frac{\cR}{6}.
    \label{eq:massshiftgeneral}
\end{equation}
In particular, the minimally coupled massless operator is reached by cancelling the link-induced curvature term with a scattering shift $\Delta M^2=-\cR/6$.  Eq.~\eqref{eq:massshiftgeneral} is the constant-curvature counterpart of the usual mass-scattering prescription; its implementation as a causal-set matrix resolvent depends on the sign convention adopted for the scattering series.
\begin{figure}[t!]
    \centering
    \subfigure[]{\includegraphics[width=0.495\textwidth]{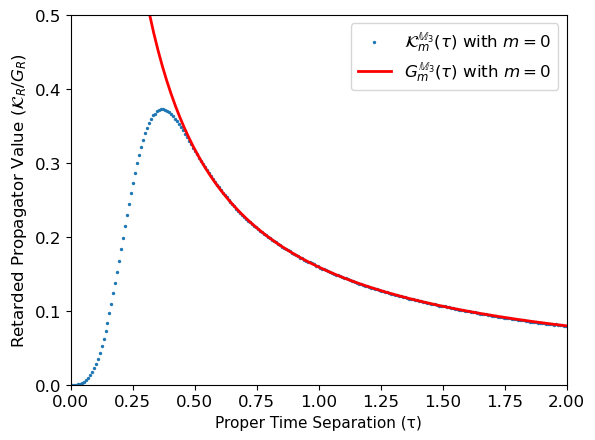}}
    \hspace{0.0001\textwidth}%
    \subfigure[]{\includegraphics[width=0.46\textwidth]{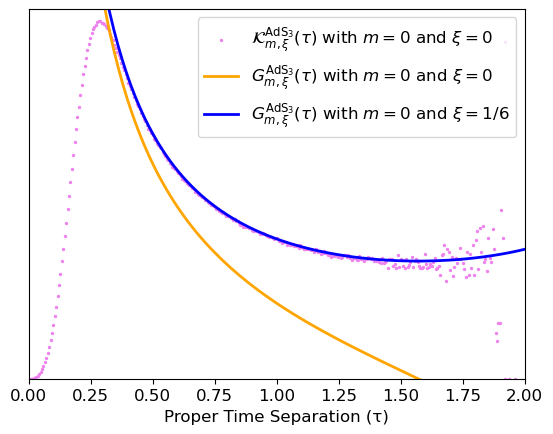}}%
   \caption{
Numerical comparison of the normalized link-square causal-set propagator $\Kcs_R$ with continuum retarded Green functions~$G_R$ in $2+1$ dimensions, using fixed density $\rho=1312$, $350$ bins and $15$ independent sprinkling averages.
(a) Minkowski spacetime, where the causal-set propagator approaches the massless continuum propagator away from the discreteness-scale region.
(b) $\mathrm{AdS}_3$, where the causal-set propagator agrees with the continuum propagator for $m=0$ and $\xi=1/6$, and deviate from the minimally coupled result with $m=0$ and $\xi=0$.
This supports the identification of the bare link-square kernel with the effective curvature coupling $\xi_{\rm link}=1/6$.
}
\label{fig:3DProps}
\end{figure}

\subsection{Numerical simulations}
\label{sec:numerics}

The preceding analysis concerns the high-density expectation value of the
link-square kernel.  We complement it with numerical sprinklings in order
to examine the approach to this limit at finite density. A fixed density $\rho=1312$, $350$ bins over the geodesic distance $\tau$ and $15$ independent sprinkling averages were used.

Fig.~\ref{fig:3DProps} (a) shows the normalized link-square kernel in
$2+1$-dimensional Minkowski spacetime.  Away from the discreteness-scale
region near $\tau=0$, the causal-set data approach the continuum massless
retarded Green function shown in Eq.~\eqref{eq:main_M}
\begin{equation}
    G^{\mathbb M_3}_{0,0}(\tau)
    =\frac{1}{2\pi\tau}. \notag
\end{equation}
The suppression at very small proper-time separation is a finite-density
effect, since the singular continuum behavior cannot be resolved below
the characteristic sprinkling scale.

The constant-curvature comparison is more discriminating.
Fig.~\ref{fig:3DProps} (b) compares the normalized bare link-square
kernel in $\AdS_3$ with both the minimally coupled massless continuum
propagator and the curvature-coupled propagator identified in
Sec.~\ref{sec:KG}.  The data clearly depart from Eq. \eqref{eq:AdSminimal}
\begin{equation}
    G^{\AdS_3}_{0,0}(\tau)
    =\frac{1}{2\pi\ell}
      \cot\left(\tau/\ell\right), \notag
\end{equation}
whereas over the resolved timelike range they closely follow Eq. \eqref{eq:main_AdS}
\begin{equation}
    G^{\AdS_3}_{0,1/6}(\tau)
    =\frac{1}{2\pi\ell} \frac{1}{\sin(\tau/\ell)}. \notag
\end{equation}
No additional mass parameter is fitted in this comparison: the effective
Klein--Gordon parameter $M^2=-1/\ell^2$ is fixed by the analytically
derived curvature coupling $\xi_{\rm link}=1/6$.
The numerical result therefore provides an independent finite-density
check of the continuum identification obtained from the high-density
asymptotics.

\subsection{Relation to link-based propagators in other dimensions}
\label{sec:dimensions}

The value $1/6$ is notable because link-centered constructions now display the following pattern across spacetime dimensions two, three, and four (summarized in Tab. \ref{tab:dimensions}):

\begin{table}[t!]
\centering
\renewcommand{\arraystretch}{1.35}
\setlength{\tabcolsep}{7pt}
\begin{tabular}{@{}lcccc@{}}
\toprule
$D$
& proportionality constant
& density-scaled link structure
& $\xi_{\rm link}$
& $\xi_{\rm conf}^{(D)}$
\\
\midrule

2
& $\frac{1}{2}e^{\gamma_E}$
& $e^{\LLL}$
& $\frac{1}{6}$
& $0$
\\[0.4em]

3
& $\left[
\frac{3}{2\pi\Gamma(2/3)}
\left(\frac{\pi}{12}\right)^{2/3}
\right]^2$
& $\rho^{1/3}\LLL^2$
& $\frac{1}{6}$
& $\frac{1}{8}$
\\[0.8em]

4
& $\frac{1}{2\pi\sqrt{6}}$
& $\rho^{1/2}\LLL$
& $\frac{1}{6}$
& $\frac{1}{6}$
\\
\bottomrule
\end{tabular}
\caption{Link-based massless propagator structures and curvature couplings.  The status of the curvature statement is not identical in the three rows: the $D=2$ result is derived perturbatively in curvature and supported numerically, the $D=3$ entry is the high-density constant-curvature result of the present work, and the $D=4$ entry follows from the conformally coupled link prescription in conformally flat de Sitter and anti-de Sitter settings.}
\label{tab:dimensions}
\end{table}
The $D=2$ entry refers to the link-exponential construction in flat space and its constant-curvature response ~\cite{HinrichsenKastrati2026,Kastrati2026Curvature}.  The $D=4$ entry is the established link prescription whose curved high-density mean gives the massless conformally coupled Green function in de Sitter spacetime and in a conformally flat anti-de Sitter patch ~\cite{AhmedDowkerSurya2017}.  The present $D=3$ result provides the intermediate-dimensional link-square member of this sequence. 

The conformal scalar coupling in $D$ spacetime dimensions is
\begin{equation}
    \xi_{\rm conf}^{(D)}=\frac{D-2}{4(D-1)}.
    \label{eq:conformal_coupling}
\end{equation}
Consequently, the equality $\xi_{\rm link}=1/6$ cannot generally be explained as conformal coupling: it is conformal only in $D=4$, whereas $\xi_{\rm conf}^{(2)}=0$ and $\xi_{\rm conf}^{(3)}=1/8$ ~\cite{BirrellDavies1982,ParkerToms2009}.  The recurrence of the same numerical coefficient in three different link-centered constructions therefore deserves a different interpretation, although the existing results do not yet constitute a proof of universality.

A suggestive continuum comparison is provided by the local heat-kernel or adiabatic expansion of a scalar Green function.  For the standard scalar Laplace-type convention, the first local curvature coefficient contains
\begin{equation}
    a_1=\left(\frac16-\xi\right)\cR,
    \label{eq:a1}
\end{equation}
so that the local scalar-curvature term vanishes at $\xi=1/6$ independently of spacetime dimension ~\cite{Vassilevich2003,ParkerToms2009}.  This $1/6$ is therefore conceptually distinct from Eq.~\eqref{eq:conformal_coupling}.  The comparison is especially relevant for links because their Poisson probability is controlled by small Alexandrov volumes, and the high-density evaluation of $\LLL^2$ is dominated by subintervals whose proper times scale to zero.  In the present $D=3$ calculation, $u,v\sim\rho^{-2/3}$ and hence the endpoint proper times satisfy $\tau_i\sim\rho^{-1/3}$.  The link-square limit is thus governed by an ultraviolet, near-null region even when the external separation $\tau$ is finite.

We stress the status of this observation.  Eq.~\eqref{eq:a1} does not derive Eq.~\eqref{eq:xilink}, nor does $\xi=1/6$ imply that curvature disappears from the full Green function.  Geometrical information remains in focusing factors, interval volumes, and higher local coefficients.  Rather, the dimension-independent zero of the first scalar curvature coefficient provides a continuum analogue of the dimension-independent value suggested by the link data.  It is therefore reasonable to ask whether $1/6$ is a characteristic ultraviolet property of propagators constructed directly from links, rather than a conformal property of any particular dimension.

This distinction also separates the present approach from causal-set d'Alembertian constructions.  Discrete retarded d'Alembertians and their curved limits provide another important route to scalar dynamics ~\cite{BenincasaDowker2010,DowkerGlaser2013,Aslanbeigi2014}, but their curvature terms are properties of those operators and need not coincide with the effective coupling extracted from a link-based propagator.  The present claim is therefore restricted to the class of link-centered kernels discussed above.

\section{Discussion and outlook}
\label{sec:outlook}

We have evaluated the mean of a simple causal-set object,
\begin{equation}
    \mathcal{N}_3\rho^{1/3}\LLL^2,
    \label{eq:finalobject}
\end{equation}
and shown that its high-density expectation reproduces a well-defined retarded scalar kernel in $2+1$-dimensional Minkowski, anti-de Sitter, and de Sitter spacetimes.  The derivation uses only Poisson link probabilities, continuum interval volumes, and a one-intermediate-point integral.  No discrete proper-time estimator or longest-chain observable is required.  This distinguishes the construction from earlier $2+1$-dimensional proposals based on Alexandrov volume or longest-chain length ~\cite{Johnston2010,AhmedDowkerSurya2017} while retaining the link matrix as the same microscopic object that already plays a central role in dimensions two and four.

The constant-curvature calculation has two structural features.  First, the leading normalization is identical to the flat normalization.  This follows from the universal short-distance form
\begin{equation}
    V(\tau_i)=\frac{\pi}{12}\tau_i^3+O(\cR\tau_i^5)
    \label{eq:universalshortV}
\end{equation}
for the small endpoint subintervals that dominate the high-density integral.  The universality of the leading small-diamond geometry is consistent with the general curved-space expansion of Alexandrov volumes ~\cite{GibbonsSolodukhin2007}.  Second, the finite external separation retains the background curvature and yields the exact direct kernels \eqref{eq:AdSlimit} and \eqref{eq:dSlimit}.  Their Klein--Gordon interpretation is $M^2_{\rm bare}=\cR/6$, or $\xi_{\rm link}=1/6$ at nominal mass zero.

The recurrence of $1/6$ in the link-exponential $D=2$ construction, the present link-square $D=3$ construction, and the link $D=4$ construction is suggestive but should not be overstated. The three results arise from distinct constructions. In particular, the four-dimensional value is also the conformal coupling, while the two- and three-dimensional values are not. The common feature is instead microscopic: all three constructions are built from links, whose defining emptiness condition probes small causal intervals. The dimension-independent heat-kernel combination $(1/6-\xi)\cR$ provides a natural continuum clue that a link-based ultraviolet mechanism may be at work.

There is also an important methodological consequence.  The massless link-centered kernels in spacetime dimensions $D=2,3,4$ now take the schematic forms
\begin{equation}
    D=2:\ e^\LLL,
    \qquad
    D=3:\ \LLL^2,
    \qquad
    D=4:\ \LLL,
    \label{eq:linkpattern}
\end{equation}
with dimension-dependent density normalizations and different ways in which path concatenation enters.  At present these forms are understood through dimension-specific constructions. The next natural step is therefore to establish a general analytic formalism that derives the appropriate function of \(\LLL\), its density scaling, its normalization, and its curvature response. The Riesz-distribution structure of these link-based constructions suggests a route toward such a general formalism, which we leave for future work. Riesz distributions, together with Laplace-transform methods, were previously used by Johnston to obtain the large-density expansion and finite-density correction terms for causal-set retarded propagators and d'Alembertian operators \cite{Johnston_2015}. Such a framework should explain, at least in the high-density continuum limit, why a single link suffices in four dimensions, why two-link paths are selected in three dimensions, why an all-link-path sum appears in two dimensions, and how these structures extend to general spacetime dimension and nonzero curvature. The results obtained here provide a concrete intermediate-dimensional test that such a formalism must reproduce.

\appendix

\section{Flat-space boundary-layer calculation}
\label{app:flat}

We derive Eq.~\eqref{eq:flat_F2_asymptotic}. The waist of $I(p,q)$ is $t=0,\,r=a$.  Define
\begin{equation}
    u=a-t-r,\qquad v=a+t-r,
    \label{eq:flatuv}
\end{equation}
so that
\begin{equation}
    t=\frac{v-u}{2},\qquad
    r=a-\frac{u+v}{2},\qquad
    \dd t\,\dd r=\frac12\dd u\,\dd v.
    \label{eq:flatjac}
\end{equation}
Near $u=v=0$, angular integration gives
\begin{equation}
    \dd V\simeq \pi a\,\dd u\,\dd v.
    \label{eq:flatwaistmeasure}
\end{equation}
The proper-time squares from the intermediate point to the endpoints are
\begin{align}
    \tau_1^2&=(a-t)^2-r^2
      =(a-t-r)(a-t+r)
      \simeq 2a\,u,
    \label{eq:flattau1}\\
    \tau_2^2&=(a+t)^2-r^2
      \simeq 2a\,v.
    \label{eq:flattau2}
\end{align}
Using $V(\tau_i)=\frac{\pi}{12}\tau_i^3$, define
\begin{equation}
    c_0(a):=\frac{\pi}{12}(2a)^{3/2}.
    \label{eq:c0}
\end{equation}
Then
\begin{equation}
    V(q,z)\simeq c_0(a)u^{3/2},\qquad
    V(z,p)\simeq c_0(a)v^{3/2}.
    \label{eq:flatVuv}
\end{equation}
Eq.~\eqref{eq:F2general} becomes
\begin{equation}
    F_2^{\Mink_3}(\tau)
    \simeq
    \rho\pi a
    \left[
       \int_0^\infty \dd u\,
       e^{-\rho c_0(a)u^{3/2}}
    \right]^2.
    \label{eq:flatfactorized}
\end{equation}
The upper limits may be extended to infinity because the boundary layer has $u,v=O(\rho^{-2/3})$.  Setting $s=\rho c_0u^{3/2}$,
\begin{equation}
    \int_0^\infty \dd u\,e^{-\rho c_0u^{3/2}}
    =\frac23\Gamma\!\left(\frac23\right)(\rho c_0)^{-2/3}.
    \label{eq:masterlaplace}
\end{equation}
Therefore
\begin{align}
    F_2^{\Mink_3}(\tau)
    &\simeq
    \frac{4\pi a}{9}\Gamma\!\left(\frac23\right)^2
    c_0(a)^{-4/3}\rho^{-1/3}
    \nonumber\\
    &=\frac{2\pi}{9\tau}\Gamma\!\left(\frac23\right)^2
    \left(\frac{\pi}{12}\right)^{-4/3}\rho^{-1/3},
    \label{eq:flatappendixresult}
\end{align}
which is Eq.~\eqref{eq:flat_F2_asymptotic}.  Multiplication by Eq.~\eqref{eq:A3} squared gives Eq.~\eqref{eq:flat_limit}.

\section{$\AdS_3$ interval volume and high-density asymptotics}
\label{app:AdS}

\subsection{Invariant distance and exact Alexandrov volume}

For two points in the coordinates of Eq.~\eqref{eq:AdSmetric}, define
\begin{equation}
    Z(x,x')=
    \frac{\cos(t-t')-\sin\chi\sin\chi'\cos(\phi-\phi')}
         {\cos\chi\cos\chi'}.
    \label{eq:AdSinvariant}
\end{equation}
Within the direct timelike normal region,
\begin{equation}
    \tau(x,x')=\ell\arccos Z(x,x').
    \label{eq:AdStauZ}
\end{equation}
For the symmetric endpoints \eqref{eq:AdSendpoints}, the causal interval is
\begin{equation}
    -a<t<a,\qquad 0<\chi<a-|t|.
    \label{eq:AdSregion}
\end{equation}
Its volume is
\begin{align}
    V_{\AdS_3}
    &=2\pi\ell^3\int_{-a}^{a}\dd t
      \int_0^{a-|t|}\dd\chi\,
      \frac{\sin\chi}{\cos^3\chi}
    \nonumber\\
    &=2\pi\ell^3(\tan a-a),
    \label{eq:AdSvolumea}
\end{align}
which becomes Eq.~\eqref{eq:AdSvolume} after $a=\tau/(2\ell)$.

\subsection{Localization near the waist}

With $u,v$ defined by Eq.~\eqref{eq:AdSuv},
\begin{equation}
    t=\frac{v-u}{2},\qquad
    \chi=a-\frac{u+v}{2},\qquad
    \dd t\,\dd\chi=\frac12\dd u\,\dd v.
    \label{eq:AdSjac}
\end{equation}
At fixed $a$ and large $\rho$, the dominant region satisfies $u,v\to0$.  Therefore
\begin{equation}
    \ell^3\frac{\sin\chi}{\cos^3\chi}\,
    \dd t\,\dd\chi\,\dd\phi
    \simeq
    \frac{\ell^3}{2}\frac{\sin a}{\cos^3a}
    \dd u\,\dd v\,\dd\phi,
    \label{eq:AdSmeasurelocal}
\end{equation}
which yields the first relation in Eq.~\eqref{eq:AdSboundarydata} after integrating $\phi$.

For $q=(a,0)$ and $z=(t,\chi,\phi)$, Eq.~\eqref{eq:AdSinvariant} reduces to
\begin{equation}
    Z(q,z)=\frac{\cos(a-t)}{\cos\chi}.
    \label{eq:AdSqzZ}
\end{equation}
Since $\chi=a-t-u$ near the past light cone of $q$,
\begin{equation}
    Z(q,z)=1-u\tan(a-t)+O(u^2).
    \label{eq:AdSZexpand}
\end{equation}
Using
\begin{equation}
    Z(q,z)=\cos(\tau_1/\ell)
    =1-\frac{\tau_1^2}{2\ell^2}+O(\tau_1^4/\ell^4),
    \label{eq:AdSsmalltauZ}
\end{equation}
we obtain
\begin{equation}
    \tau_1^2=2\ell^2\tan(a-t)u+O(u^2)
    \simeq2\ell^2\tan a\,u.
    \label{eq:AdStau1local}
\end{equation}
The same calculation near the future light cone of $p$ gives
\begin{equation}
    \tau_2^2\simeq2\ell^2\tan a\,v.
    \label{eq:AdStau2local}
\end{equation}
Because these endpoint intervals shrink as $\rho\to\infty$, only their leading flat volume is needed:
\begin{equation}
    V(\tau_i)=\frac{\pi}{12}\tau_i^3
    +O(\tau_i^5/\ell^2).
    \label{eq:AdSlocalsmallV}
\end{equation}
Eqs.~\eqref{eq:AdStau1local}--\eqref{eq:AdSlocalsmallV} give Eq.~\eqref{eq:AdSboundarydata} and Eq.~\eqref{eq:ca}.

Finally,
\begin{align}
    F_2^{\AdS_3}
    &\simeq
    \rho\pi\ell^3\frac{\sin a}{\cos^3a}
    \int_0^\infty\dd u\int_0^\infty\dd v\,
    e^{-\rho c(a)u^{3/2}}e^{-\rho c(a)v^{3/2}}
    \nonumber\\
    &=\rho\pi\ell^3\frac{\sin a}{\cos^3a}
      \left[
      \frac23\Gamma\!\left(\frac23\right)
      (\rho c(a))^{-2/3}
      \right]^2,
    \label{eq:AdSfullasymptotic}
\end{align}
from which Eqs.~\eqref{eq:AdSF2asymptotic} and \eqref{eq:AdSlimit} follow.

\section{$\dS_3$ interval volume and high-density asymptotics}
\label{app:dS}

\subsection{Exact interval volume}

The $\dS_3$ invariant in the coordinates of Eq.~\eqref{eq:dSmetric} is
\begin{equation}
    Z(x,x')=
    \frac{\cos\Theta_{S^2}-\sin t\sin t'}{\cos t\cos t'},
    \label{eq:dSinvariant}
\end{equation}
where
\begin{equation}
    \cos\Theta_{S^2}
    =\cos\chi\cos\chi'
     +\sin\chi\sin\chi'\cos(\phi-\phi').
    \label{eq:S2angle}
\end{equation}
For timelike separation, $Z=\cosh(\tau/\ell)$.  Along the central geodesic $\chi=0$,
\begin{equation}
    \tau
    =\ell\int_{-a}^{a}\frac{\dd t}{\cos t}
    =2\ell\operatorname{artanh}(\sin a),
    \label{eq:dScentralproper}
\end{equation}
which is Eq.~\eqref{eq:dStauofa}.

The conformal coordinates preserve the null boundaries, so the symmetric Alexandrov interval again satisfies Eq.~\eqref{eq:AdSregion}.  Its volume is now
\begin{align}
    V_{\dS_3}
    &=2\pi\ell^3\int_{-a}^{a}\frac{\dd t}{\cos^3t}
      \int_0^{a-|t|}\sin\chi\,\dd\chi
    \nonumber\\
    &=4\pi\ell^3\int_0^a\frac{1-\cos(a-t)}{\cos^3t}\,\dd t
    \nonumber\\
    &=2\pi\ell^3\left[
       \operatorname{artanh}(\sin a)-\sin a
    \right].
    \label{eq:dSvolumea}
\end{align}
Writing $b=\tau/(2\ell)=\operatorname{artanh}(\sin a)$ implies $\sin a=\tanh b$, and Eq.~\eqref{eq:dSvolume} follows.

\subsection{Waist expansion}

Use again
\begin{equation}
    u=a-t-\chi,\qquad v=a+t-\chi.
    \label{eq:dSuvagain}
\end{equation}
At the waist $t=0$, $\chi=a$, Eq.~\eqref{eq:dSmeasureR} gives
\begin{equation}
    \dd V\simeq
    \frac{\ell^3\sin a}{2}\,
    \dd u\,\dd v\,\dd\phi
    \longrightarrow
    \pi\ell^3\sin a\,\dd u\,\dd v
    \label{eq:dSmeasureapp}
\end{equation}
when the angular integral is performed.

For the later endpoint $q=(a,0)$,
\begin{equation}
    Z(q,z)=
    \frac{\cos\chi-\sin a\sin t}{\cos a\cos t}.
    \label{eq:dSqzZ}
\end{equation}
On the past null cone of $q$, $\chi=a-t$ and $Z=1$.  Setting $\chi=a-t-u$ gives
\begin{equation}
    Z(q,z)
    =1+u\left(\tan a-\tan t\right)+O(u^2).
    \label{eq:dSZexpand}
\end{equation}
Near the waist, $t\to0$, and comparison with
\begin{equation}
    \cosh(\tau_1/\ell)
    =1+\frac{\tau_1^2}{2\ell^2}+O(\tau_1^4/\ell^4)
    \label{eq:dScoshsmall}
\end{equation}
therefore yields
\begin{equation}
    \tau_1^2\simeq2\ell^2\tan a\,u,
    \qquad
    \tau_2^2\simeq2\ell^2\tan a\,v.
    \label{eq:dStau12}
\end{equation}
The leading small-subinterval volume is again $\pi\tau_i^3/12$, which establishes Eq.~\eqref{eq:dSsmallV} with the same $c(a)$ as in Eq.~\eqref{eq:ca}.  Consequently,
\begin{align}
    F_2^{\dS_3}
    &\simeq
    \rho\pi\ell^3\sin a
    \left[
       \frac23\Gamma\!\left(\frac23\right)
       (\rho c(a))^{-2/3}
    \right]^2.
    \label{eq:dSfullasymptotic}
\end{align}
After multiplication by $\mathcal{N}_3\rho^{1/3}$,
\begin{equation}
    \Kcs^{\dS_3}_{0}
    =\frac{1}{4\pi\ell}\frac{\cos^2a}{\sin a}.
    \label{eq:dSinA}
\end{equation}
Since $\sin a=\tanh b$ and $b=\tau/(2\ell)$,
\begin{equation}
    \frac{\cos^2a}{\sin a}
    =\frac{\operatorname{sech}^2b}{\tanh b}
    =\frac{2}{\sinh(2b)},
    \label{eq:dSidentity}
\end{equation}
which shows Eq.~\eqref{eq:dSlimit}.


\begin{acknowledgments}
This work was supported financially by the Employment Agency Bonn and the University of Würzburg.
The author thanks H. Hinrichsen for critical reading of the manuscript and David A. Meyer for useful discussions.
\end{acknowledgments}

\bibliography{references}

\end{document}